\documentclass[letterpaper,twocolumn,10pt]{article}
\usepackage{usenix2019_new,epsfig,endnotes}
\usepackage{times}
\usepackage{hyperref}
\usepackage{url}     

\usepackage{booktabs} 
\usepackage{balance}
\usepackage{float}
\usepackage{subfig} 
\usepackage{graphicx}
\usepackage{listings}
\usepackage{setspace}
\usepackage{xspace}
\usepackage{multirow}

\usepackage[font={footnotesize,it}]{caption}
\usepackage[export]{adjustbox}
\usepackage[skip=1ex]{caption}

\usepackage{amssymb}
\usepackage{pifont}

\usepackage{tikz}

\newcommand{\figref}[1]{{Figure~\ref{#1}}}
\newcommand{\secref}[1]{{\S\ref{#1}}}
\newcommand{\thmref}[1]{{Theorem~\ref{#1}}}

\newtheorem{sectheorem}{Theorem}[subsection]

\newcommand{\tabref}[1]{{Table~\ref{#1}}}

\newcommand{\cmark}{\ding{51}}%

\newcommand{\JK}[1]{}
\newcommand{\CR}[1]{}
\renewcommand{\CR}[1]{{\color{black}{#1}}}

\newcommand{\junaid}[1]{}
\newcommand{\aditya}[1]{}
\renewcommand{\JK}[1]{{\color{black}{#1}}}
\renewcommand{\aditya}[1]{{\color{blue}{\bf AA:#1}}}
\renewcommand{\junaid}[1]{{\color{orange}{\bf JK:#1}}}

\newcommand{\compactcaption}[1]{\vspace{-0.3em}\caption{#1}\vspace{-1.5em}}

\newcommand{\name}{CHC\xspace}

\definecolor{nodecolor}{RGB}{255,115,115}

\newcommand*\circled[1]{\tikz[baseline=(char.base)]{
		\node[shape=circle,draw=nodecolor,inner sep=1pt, fill=nodecolor] (char) {#1};}}

\begin{document}
	\font\titlefont=cmr10 at 20pt
	\title{Correctness and Performance for Stateful Chained Network Functions}


	\author{
	{\rm Junaid Khalid \hspace{0.1in} Aditya Akella}\\
	{University of Wisconsin - Madison}
	}

	\maketitle


\noindent
{\bf Abstract:} Network functions virtualization (NFV) allows
operators to employ NF chains to realize custom policies, and
dynamically add instances to meet demand or for failover. NFs maintain
detailed per- and cross-flow state which needs careful management,
especially during dynamic actions. Crucially, state management must:
(1) ensure NF {\em chain-wide} correctness and (2) have good
performance. To this end, we built
\name, an NFV framework that leverages an external state store
coupled with state management algorithms and metadata maintenance for
correct operation even under a range of failures. Our evaluation shows
that \name can support $\sim$10Gbps per-NF throughput and $<0.6\mu$s increase in \CR{median} per-NF packet processing latency, and chain-wide correctness at little additional cost.



	\vspace*{-.1in}
\section{Introduction}
\label{sec:intro}


NFV vastly improves network management. It allows operators to
implement rich security and access control policies using NF
chains~\cite{nfvwhitepaper,flowtags,boucadair2013differentiated,etsi,nsh}. Operators
can overcome NF failure and performance issues by spinning up
additional instances, and dynamically redistributing
traffic~\cite{gember2013stratos,ftmb}.

\begin{figure}[t!]
	\centering 
	\subfloat[][]{%
		\label{fig-physical-chain}	
		\includegraphics[scale=0.65]{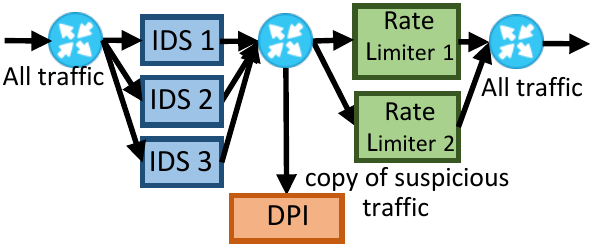}%
	}
	\subfloat[][]{%
		\label{fig-example-topo}
		\label{fig-logical-chain}		
		\includegraphics[scale=0.65]{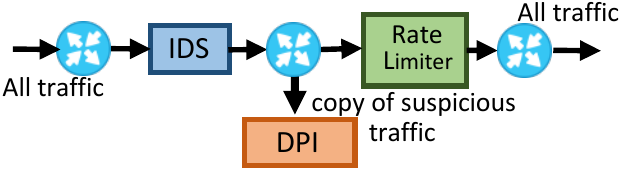}%
		
	}

	\compactcaption{(a) Example NFV chain with many instances per NF (b) logical view with infinite capacity NFs/links for COE.}
	\label{fig-nfv-chain}
	\vspace{0.3em}
\end{figure}

To be applicable to enforcing policies correctly, NFV must provide
{\em chain output equivalence} (COE): given an input packet stream, at
any point in time, the collective action taken by all NF instances in
an NFV chain (\figref{fig-physical-chain}) must match that taken by an
hypothetical equivalent chain with infinite capacity always available single NFs (\figref{fig-logical-chain}). COE must hold under dynamics:
under NF instance failures/slowdowns, traffic reallocation for load
balancing/elastic scaling, etc. Given that NFV is targeted for cloud
and ISP deployments, COE should not come at the cost of {\em
performance}.

These goals are made challenging by NFs' {\em statefulness}.  Most NFs
maintain detailed internal state that could be updated as often as per
packet. Some of the state may be shared across instances. For example,
the IDS instances in Figure~\ref{fig-physical-chain} may share {\em
  cross-flow state}, e.g., per port counters. They may also maintain
{\em per-flow} state, e.g., bytes per flow, which is confined to
within an instance.

Ensuring COE under statefulness requires that, as traffic is being
processed by many instances, or being reassigned across instances,
updates to state at various NFs must happen in a ``correct''
fashion. For example, shared state updates due to packets arriving at
IDS1 must be reflected at IDS2; likewise, when reallocating a flow,
say $f_1$, from IDS1 to 2, $f_1$'s state should be updated due to
in-flight $f_1$ packets arriving at both IDSes 1 and 2. Finally, {\em
  how} the state is updated can determine an NF's action. For example,
the off-path Trojan detector~\cite{lorenzos_work} in Figure~\ref{fig-example-trojan} relies
on knowing the exact order in which connection attempts were
made. When there is a discrepancy in the order observed w.r.t. the
true order -- e.g., due to intervening NFs running slow or failing --
the Trojan detector can arrive at incorrect decisions, violating COE.

Many NFV frameworks exist
today~\cite{ftmb,pico,splitmerge,E2,OpenBox,statelessnf,S6}. Several of
them focus on managing NF state migration or updates upon traffic
reallocation during scaling or
failover~\cite{ftmb,pico,splitmerge,opennf,S6}. However, they either violate COE, or suffer from poor performance (or both).

First, most systems {\bf ignore shared
  state}~\cite{ftmb,pico,splitmerge,E2}.  They assume
that NFs do not use cross-flow state, or that traffic can be split
across NF instances such that sharing is completely
avoided. Unfortunately, neither assumption is valid; many
NFs~\cite{broids,squid,re,prads} have cross-flow state, and the need
for fine-grained traffic partitioning for load balancing can easily
force cross-flow state sharing across instances. Because shared state is
critical to NF processing, ignoring how it is updated can lead to
inconsistent NF actions under dynamics, violating COE (\secref{today}).

Second, existing approaches {\bf cannot support chain-level
  consistency}. They cannot ensure that the order of updates made to
  an NF's state (e.g., at the Trojan detector~\cite{lorenzos_work}
  in~\figref{fig-example-trojan}) are consistent with the input packet
  stream. This inability can lead to NFs arriving at incorrect
  decisions, e.g., missing out on detecting attacks (as is the case
  in~\figref{fig-example-trojan}), violating COE. Similar issues arise
  in the inability to correctly suppress spurious duplicate updates
  observed at an NF due to recovery actions at upstream NFs
  (\secref{requirements}).


Finally, existing frameworks impose {\bf high overhead} on state
maintenance, e.g., 100s of milliseconds to move per-flow state 
across instances when traffic is reallocated (\secref{today}).

We present a new NFV framework, \name (``correct, high-performance
chains''), which overcomes these drawbacks. For COE, \name uses three
building blocks. \name stores NF state in an in-memory
{\bf external state store}. This ensures that state continues to be
available after NF instances' recover from failure, which is necessary
for COE. Second, it maintains simple {\bf metadata}. It adds a
``root'' at the entry of a chain that: (1) applies a unique logical
clock to every packet, and (2) logs packets whose processing is still
ongoing in the chain. At the store and NFs, \name tracks packet clocks
along with update operations each NF issues. Clocks help NFs to reason
about relative packet ordering irrespective of intervening NFs'
actions, and, together with datastore logs, help suppress
duplicates. We develop failure recovery protocols which leverage
clocks and logs to ensure correct recovery from the failure. In
Appendix \ref{proofs}, we prove their correctness by showing
that the recovered state is same as if no failure has occurred,
thereby ensuring COE. 

State externalization can potentially slow down performance of state
reads/writes.  Thus, for performance, \name introduces {\bf NF-aware
algorithms} for shared state management. It uses scope-awareness of
state objects to partition traffic so as to minimize cross-instance
shared state coordination. It leverages awareness of the state access
patterns of NFs to implement strategies for shared state
caching. Because most NFs today perform a simple set of state update
operations, \name offloads operations to the state store, which
commits them in the background. This speeds up shared state updates
-- all coordination is handled by the store which
serializes the operations issued by multiple NF instances.


We built a multi-threaded C++ prototype of \name along with
four NFs. We evaluate this prototype using two campus-to-EC2
packet traces.  \JK{We find that \name's state management optimizations
reduce latency overhead to {\bf 0.02$\mu$s - 0.54$\mu$s} per packet
compared to traditional NFs (no state externalization).} \name failover offers {\bf 6X} better
75\%-ile per packet latency than~\cite{ftmb}. \name is 99\% faster in
updating strongly consistent shared state, compared to
~\cite{opennf}. \name obtains per-instance
throughput of 9.42Gbps -- {\bf same as maximum achievable} with
standalone NFs. \name's support for chain-wide guarantees adds little
overhead, but {\bf eliminates false positives/negatives} seen when
using certain security NFs in existing NFV frameworks. Thus, \name is
the only framework to support COE, and it does so at state-of-the-art
performance.



	\section{Motivation}

\label{s:mot}

NFV allows operators to connect NFs together in chains, where each
type of NF can use multiple instances to process input traffic
demand. Use of software NFs and SDN~\cite{simple} means that when
incoming traffic load spikes, or processing is unbalanced across 
instances, operators can scale up by adding NF instances and/or
reallocate flow processing  across
instances. Furthermore, hot-standby NFs can be used to continue packet
processing when an instance crashes. Due to these benefits, cloud
providers and ISPs are increasingly considering deploying NFV in their
networks~\cite{etsiwhitepaper}.


\subsection{Key Requirements for COE}
\label{requirements}

NFV chains are central to security and compliance policies,
they must always operate correctly, i.e., ensure COE
(\secref{sec:intro}).  Ensuring COE is challenging: (1) NFs are stateful; they maintain state objects for individual and group of flows. These state objects may be updated on every packet and the value of these state objects may be used to determine the action on the packet. This requires support for fine gained NF state management. (2) In addition to this, COE also require that the per-NF and chain-wide state updates are consistent with the input packet stream.
(3) Since chaining may create a dependency between the action taken in upstream instances and its downstream instances, it is important that the impact of a local action taken for failure recovery should be isolated from the rest of the chain. These challenges naturally map to three classes of requirements for supporting COE:

\vspace{0.05in}
\noindent {\bf State Access:} The processing of each packet requires
access to up-to-date state; thus, the following requirement are 
  necessary to ensure COE under dynamics:

$\bullet$ (R1) {\em State availability:} When an NF instance fails, all state it
has built up internally disappears. For a failover instance to take
over packet processing it needs access to the state that the failed
instance maintained just prior to crashing.

$\bullet$ (R2) {\em Safe cross-instance state transfers:} When traffic is
reallocated across NF instances to rebalance load, the state
corresponding to the reallocated traffic (which exists at the old
instance where traffic was being processed) must be made available at
the reallocated traffic's new location.

\vspace{0.05in}\noindent {\bf Consistency:} Action taken by a given NF
instance may depend on shared-state updates made by other instances of
the same NF, or state actions at upstream NFs in the chain. Ensuring
that said NF instances' actions adhere to COE boils down to following
requirements:





$\bullet$ (R3) {\em Consistent shared state:} Depending on the nature
of an NF's state, it may not be possible to completely avoid sharing a
subset of it across instances, no matter how traffic is partitioned
(e.g., port counts at the IDSes in~\figref{fig-physical-chain}). Such
state needs to be kept consistent across the instances that are
sharing; that is, writes/updates made locally to shared state by
different instances should be executed at all other instances sharing
the state in the same global order. Otherwise, instances may end up
with different views of shared state leading to inconsistent and hence
incorrect actions.
\begin{figure}[t]
	
	
	\centering
	\includegraphics[width=0.29\textwidth]{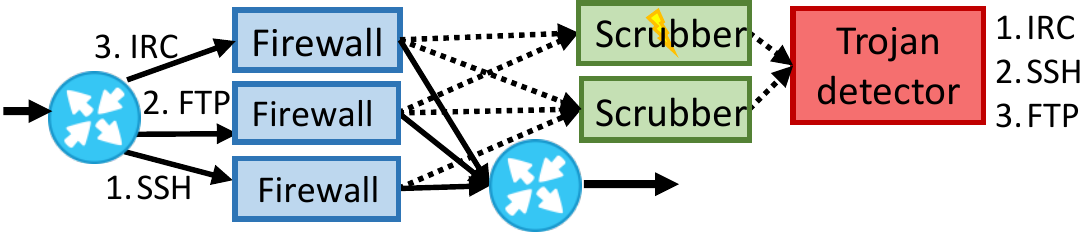}
	\compactcaption{Illustrating violation of chain-wide ordering.}
	\label{fig-example-trojan}
	
	\vspace{0.2em}
\end{figure}

$\bullet$ (R4) {\em Chain-wide ordering:} Some NFs rely on knowing the order in
which traffic entered the network.
Consider~\figref{fig-example-trojan}. The off-path Trojan
detector~\cite{lorenzos_work} works on a copy of traffic and
identifies a Trojan by looking for this sequence of steps:
(1) open an SSH connection; (2) download HTML, ZIP, and EXE files over
an FTP connection; (3) generate IRC activity. When a Trojan is
detected, the network blocks the relevant external host. A different
order does not necessarily indicate a Trojan. It is crucial that the
Trojan detector be able to reason about the true arrival order as seen
at traffic input.

In~\figref{fig-example-trojan}, either due to one of the scrubbers
being slowed down due to resource contention or recovering from
failure~\cite{ftmb}, the order of connections seen at the Trojan
detector may differ from that in the traffic arriving at the input
switch.  Thus, the Trojan detector can either incorrectly mark Trojan
traffic as benign, or vice versa.  When multiple
  instances of the Trojan detector are used, the problem is compounded
  because it might not be possible to partition traffic such that all
  three flows are processed at one instance.


$\bullet$ (R5) {\em Duplicate suppression:} In order to manage
straggler NFs, NFV frameworks can (a) deploy clones initialized with the
state of a slow NF instance; (b) use packet replay to bring the clone up
to speed with the straggler's state since state initialization; and (c) replicate packets to the straggler and clone (\secref{straggler}). Depending on when the
clone's state was initialized, replay can lead to duplicate state
updates at the straggler. Also, the original and clone instances will
then both generate duplicate output traffic. Unless such duplicate
updates and traffic are suppressed, the actions of the straggler and
of downstream NFs can be impacted (spurious duplicates may trigger an
anomaly). The need for duplicate suppression also arises during fault
recovery (\secref{s:ft-in}).

\vspace{0.05in}
\noindent
{\bf Isolation:} NFs in a chain should not be impacted by failure
recovery of NFs upstream from them. Specifically:

$\bullet$ (R6) {\em Safe chain-wide recovery:} When NF
failures occur and recovery takes place, it is
important that the state at each NF in the
chain subsequent to recovery have the same
value as in the no-failure case. In other
words,  actions taken during
recovery should not impact the processing, state,
or decisions of NFs upstream or downstream
from the recovering NF --- we will exemplify
this shortly when we describe failings of
existing systems in meeting this
requirement.




The network today already reorders or drops packets.  Our goal is to
ensure that NF replication, chaining, and traffic reallocation
together do not induce artificial ordering or loss on top of
network-induced issues. This is particularly crucial for many
important off-path NFs (e.g., DPI engines and exfiltration checkers)
which can be thwarted by artificially induced reordering or loss.

\vspace{-.1in}
\subsection{Related work, and Our Contributions}
\label{today}

A variety of NFV frameworks exist
today~\cite{ftmb,opennf,statelessnf,pico,splitmerge,E2,pga,OpenBox,xomb,comb,flowtags,S6}. We
review their drawbacks below.

{\bf Incomplete support for correctness requirements:} Most existing
frameworks focus on handling requirements R1 and/or
R2. Split/Merge~\cite{splitmerge}, OpenNF~\cite{opennf} and S6~\cite{S6} support
cross-instance state transfers (R2).  FTMB~\cite{ftmb} and Pico
Replication~\cite{pico} focus on state availability
(R1).

More fundamentally, Split/Merge, Pico Replication and FTMB focus on
availability of the state contained entirely within an NF
instance. They either ignore state shared across instances, or focus
on the small class of NFs where such state is not used. Thus, these
frameworks cannot handle R3.

\JK{Among existing frameworks, only OpenNF and S6 can support
consistency for shared state (R3), but this comes at high performance
cost. For example, OpenNF imposes a 166$\mu$s per packet 
overhead to ensure strong consistency! (\secref{s:eval}). Similarly, S6 cannot support frequent updates to strongly consistent shared state.} 

Equally crucially, all of the above frameworks focus on a single NF; they cannot handle chains. Thus, none of them support chain-wide
ordering (R4).

\JK{Support for R5 is also missing. StatelessNF~\cite{statelessnf} and S6~\cite{S6} update shared state in an external store or remote NF, respectively,
but they do not support atomic updates to all state objects an instance
can access.} Thus, when a clone is created to mitigate a straggler
off-path NF (as outlined above), the straggler may have
updated other state objects that are not reflected in the clone's initialized state. Upon replay, the straggler can make duplicate state updates (likewise, duplicate
packets can also arise). For the same reason, R6 is also violated:
when an NF fails over, replaying packets to bring the recovery NF up to
speed can result in duplicate processing in downstream NFs. 

{\bf State management performance is poor:} FTMB's periodic
checkpointing significantly inflates NF packet processing latency
(\secref{s:eval}). As mentioned above, OpenNF imposes performance
overhead for shared state. The overhead is high even for
cross-instance transfers of per-flow state: this is because such
transfers require extracting state from an instance and installing it
in another while ensuring that incoming packets are directed to the
state's new location. 




{\bf Our contributions:} How do we support requirements R1-R6 while
ensuring good state management performance?  Some NFs or operating
scenarios may just need a subset of R1-R6. However, we seek a {\em
  single framework} that meets all requirements/scenarios because,
with NFV becoming mainstream, we believe we can no longer trade-off
general correctness requirements for performance or functionality
(specific NFs). Thus, we identify {\em basic building blocks} and
study how to synthesize them into one framework. \CR{We have set ourselves the ambitious goal of designing a single generic NFV framework to support all of these requirements, though some NFs may only need support for a subset of these requirements.} Building such a
framework is especially challenging because we must carefully deal
with shared state and NF chaining.

\begin{figure}[t!]
	\centering 
	\subfloat[][]{%
		\includegraphics[width=0.35\textwidth]{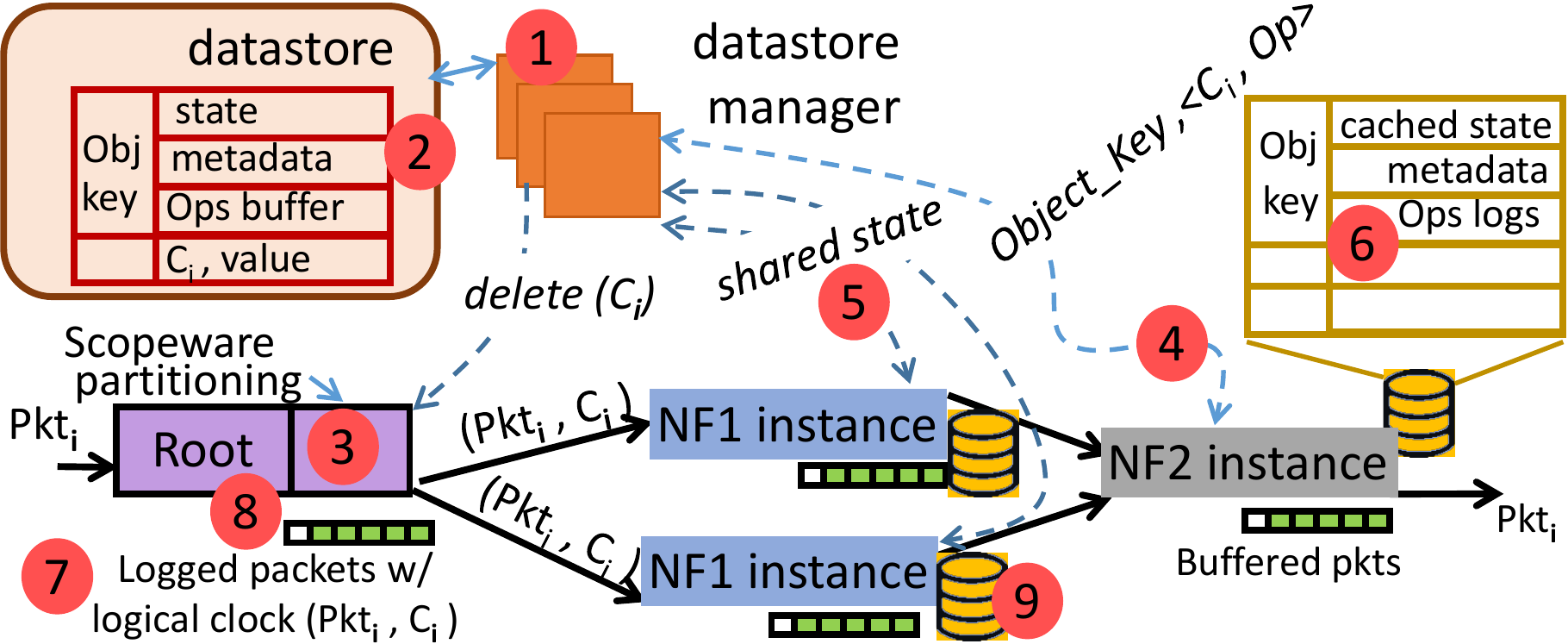}
		\label{fig-overview}
		\label{fig-dag-mngnt}
	}
	\subfloat[][]{%
		\includegraphics[width=0.07\textwidth]{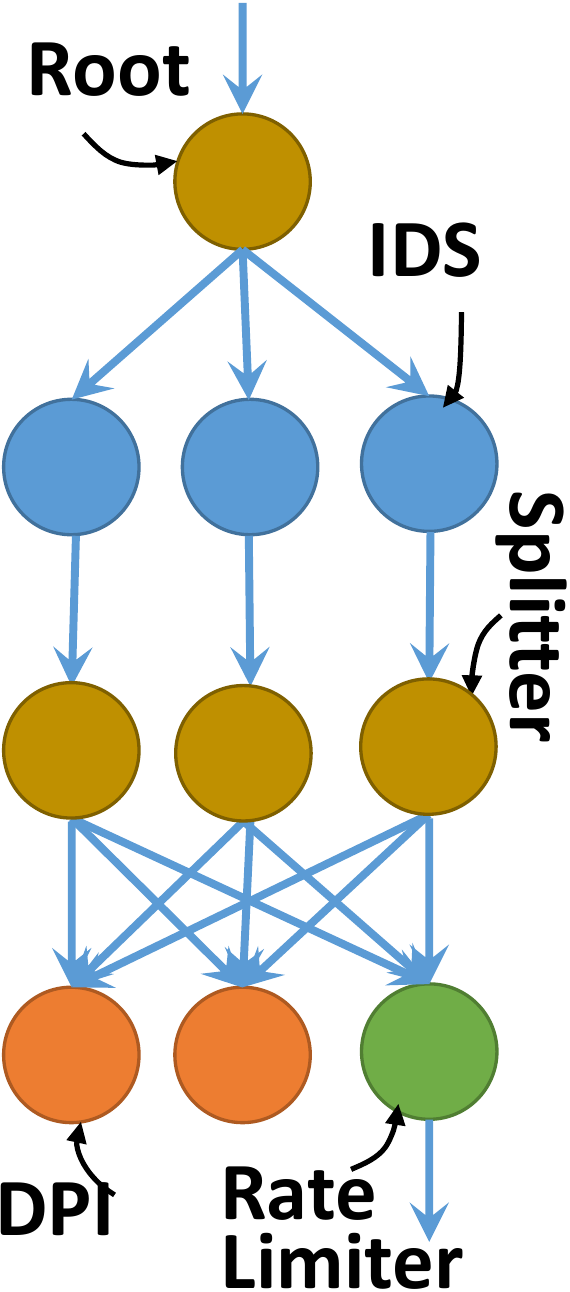}%
%
		\label{fig-dag}
		\label{fig-dag-phy} 
		
	}

		\vspace{-0.3em}
	\compactcaption{(a) \name architecture; (b) Physical chain that \name runs.}
\end{figure}

Our system, \name, has three building blocks
  (\figref{fig-overview}): We maintain NF state in a {\bf state store
    external to NFs} (\circled{1}; \secref{s:design}).  NFs access the
  store to read/write relevant state objects. This ensures state
  availability (R1). The store's state object metadata simplifies
  reasoning about state ownership and concurrency control across
  instances (\circled{2}; \secref{storage-sec}). This makes state
  transfer safety (R2) and shared state consistency (R3) simple and
  efficient (\secref{g:scaling}).


%
%
%

We propose {\bf NF state-aware algorithms} for good state read/write
performance which is a key concern with state externalization. These
include (\secref{storage-sec}): automatic state scope-aware traffic
partitioning to minimize shared-state coordination (\circled{3});
asynchronous state updates for state that is updated often but read
infrequently; this allows packet processing to progress unimpeded
(\circled{4}); NFs sending update operations, as opposed to updated
state, to the store, which simplifies synchronization and
serialization of shared-state updates (\circled{5}); scope- and access
pattern-aware state caching strategies, which balances caching
benefits against making cache updates immediately visible to other
instances (\circled{6}).

Finally, we maintain a small amount of {\bf metadata -- clocks and
  logs}. We insert per packet logical clocks (\circled{7};
\secref{runtime-management}) which directly supports cross-instance
ordering (R4). We couple clocks with logs to support duplicate
suppression (R5; \secref{straggler}) and COE under failover of NFs and
framework components (R6; \secref{s:ft-in}). We log every packet that
is currently being processed at some NF in the chain (\circled{8}).
Logged packets are replayed across the entire chain during
failover. At the state store, we store logical clocks of packets along
with the state updates they resulted in, which aids duplicate
suppression. At each NF, we store packet clocks along with the update
operations issued and the most recently read state value
(\circled{9}). Together with state store snapshots, these NF-side logs
support COE under datastore recovery.

Though StatelessNF~\cite{statelessnf} first advocated for externalizing
state, but it has serious issues. Aside from a lack of support for
R4--R6, it lacks atomic state updates: when a single NF fails after
updating some but not all state objects, a failover NF can boot up
with incorrect state!  It requires locks for shared state updates,
which degrades performance. Also, it assumes Infiniband networks for performance.



	\section{Framework: Operator View}
\label{sec:api}

In \name, operators define ``logical'' NF chains (such
as~\figref{fig-logical-chain}) using a DAG API. We elide low level
details of the API, such as how policies are specified,
and focus on aspects related to correctness and
performance. Each ``vertex'' of the DAG is an NF and consists of
operator supplied NF code, input/output, configuration, and state
objects. Edges represent the flow of data (packets and/or
contextual output). 

\label{vertman}

The \name framework compiles the logical DAG into a physical DAG with
logical vertex mapped to one or more {\em instances}
(\figref{fig-dag-phy}).  For example, the IDS
in~\figref{fig-logical-chain} is mapped to three instances
in~\figref{fig-dag-phy}.  The operator can provide default parallelism
per vertex, or this can be determined at run time using
operator-supplied logic (see below). \name deploys the instances
across a cluster.  Each instance processes a partition of the traffic
input to the logical vertex; \name automatically determines the traffic
split to ensure even load distribution (\secref{s:design}).

The \name framework supports chain elastic scaling and straggler
mitigation. Note that the logic, e.g., when to scale is not our focus;
we are interested in high performance state management and COE during
such actions. Nevertheless, we outline the operator-side view
for completeness: Operators must supply relevant logic for each vertex
(i.e., scaling\footnote{e.g., ``when input traffic volume increased
  by a certain $\theta$''}; identifying
stragglers\footnote{when an instance processing $\theta\%$ slower than
  other instances}). \name executes the logic with input from a
``vertex manager'', a logical entity is responsible for collecting
statistics from each vertex's instances, aggregating them, and
providing them periodically to the logic.

Based on user-supplied logic, \name redirects traffic to (from) scaled up
(down) NF instances or clones of straggler NFs. \name manages
state under such dynamic actions to ensure COE. \name also
ensures system-wide fault tolerance. It automatically recovers from
failures of NFs or of \name framework components while always
preserving COE.

\section{Traffic and State Management}
\label{s:design}

We discuss how \name processes traffic and manages state.  The
framework automatically partitions traffic among NF instances
(\secref{scope-aware}) and manages delivery of packets to downstream NFs (\secref{comm}). As packets flow, different NFs
process them and update state in an external store; \name leverages
several algorithms for fast state I/O; the main challenge here is
dealing with shared state (\secref{storage-sec}).
\subsection{Traffic partitioning} 
\label{scope-aware}



\name performs {\em scope-aware partitioning}: traffic from an
upstream instance is partitioned across downstream instances such
that: (1) each flow is processed at a single instance, (2) groups of
flows are allocated to instances such that most state
an instance updates for the allocated flows is not updated by
other instances, and (3) load is balanced. \#1 and \#2 reduce
the need for cross-instance coordination for shared state.



In \name, state scope is a {\em first-class entity}.  A function { \it
  .scope()} associated with a vertex program returns a list of scopes
i.e., the set of packet header fields which are used to key into the
objects that store the states for an NF; i.e., these are the different
granularities at which states can be queried/updated. \name orders the
list from the most to least fine grained scope. Suppose the DPI vertex
in \figref{fig-example-topo} has two state objects: one corresponding
to records of whether a connection is successful or not; and another
corresponding to the number of connections per host. The scope for the
former is the 5-tuple ({\it src IP, dst IP, src port, dst port,
  protocol}); the scope for the latter is {\it src IP}.

\name first attempts to partition traffic at instances immediately
upstream (which, for the DPI in~\figref{fig-example-topo} would be the
IDSes) based on the most coarse-grained state scope (for the DPI this
is {\it src IP}); such splitting results in no state sharing at the
downstream (DPI) instances. However, being coarse grained, it may
result in uneven load across instances. The framework gathers this
information via the (DPI) vertex manager. It then
considers progressively finer grained scopes and repeats the above
process until load is even. 



The final scope to partition on is provided in common to the {\em
  splitters} upstream. The framework inserts a splitter after every NF
instance (\figref{fig-dag-phy}). The splitter partitions the output
traffic of the NF instance to instances downstream.

The {\em root} of a physical DAG is a special splitter that receives
and splits input traffic. Roots can use multiple instances to handle
traffic; in \name, we fix root parallelism to some constant $R$.
Network operators are required to statically partition traffic among
the $R$ roots such that the traffic processed by a root instance has
no overlap in any of the 5-tuple dimensions with that processed by
another instance.

 \subsection{Communication}
 \label{comm}
 
 Inter-NF communication is asynchronous and non-blocking. Each NF's
 outputs are received by the \name framework which is responsible for
 routing the output to downstream instances via the splitter. The
 framework stores all the outputs received from upstream instances in
 a queue per downstream instance; downstream instances poll the queue
 for input.  This approach offers three benefits: (a) upstream
 instances can produce output independent of the consumption rate of
 downstream instances, (b) the framework can operate on queue contents
 (e.g., delete messages before they are processed downstream), which
 is useful for certain correctness properties, e.g., duplicate
 suppression (\secref{runtime-management}), (c) user logic can use
 persistent queues to identify stragglers/uneven load.
 


\subsection{State  Maintenance}
\label{storage-sec}



\name{} {\em externalizes} NF state and stores it in an external
distributed key-value datastore. Thus, state survives NF crashes,
improving availability and satisfying requirement R1
(\secref{s:mot}). All state operations are managed by the datastore
(\figref{fig-overview}). As described below, \name incorporates
 novel algorithms and metadata to improve performance (\tabref{ts:iostrategies}).

\label{metadata}









\begin{table}[]
	\centering
	\scriptsize
	\begin{tabular}{p{0.7cm}||p{1.4cm}|p{1.40cm}|p{1.3cm}|p{1.9cm}}
		Scope& Any&Per-flow&Cross-flow&Cross-flow\\
		
		\hline
		
		Access pattern&Write mostly, read rarely&Any&Write rarely (read heavy)&Write/read often\\
		
		\hline
		\hline
		
		&Non-blocking ops. No caching&Caching \textbackslash w periodic non- blocking flush&Caching \textbackslash w callbacks&Depends upon traffic split. Cache, if split allows; flush periodically \\
	\end{tabular}
	\caption{Strategies for state management performance}
	\vspace{-0.2in}
	\label{ts:iostrategies}	
\end{table}

{\bf State metadata:} The datastore's client-side library appends
metadata to the key of the state that an NF instance stores.  This
contains vertex{\_}ID and instance{\_}ID, which are immutable and are
assigned by the framework. In \name, the key for a per-flow (5 tuple)
state object is: {\it vertex{\_}ID + instance{\_}ID + obj{\_}key},
where obj{\_}key is a unique ID for the state object. The
instance{\_}ID ensures that only the instance to which the flow is
assigned can update the corresponding state object. Thus, this
metadata simplifies reasoning about ownership and concurrency
control. Likewise, the key for shared objects, e.g., {\it
  pkt{\_}count}, is: {\it vertex{\_}ID + obj{\_}key}. All the instances
of a logical vertex can update such objects. When two logical vertices
use the same key to store their state, vertex{\_}ID prevents any
conflicts.


  {\bf Offloading operations:} Most NFs today perform simple
  operations on state. \tabref{t:operations} shows common examples. In
  \name, an instance can {\em offload operations} and instruct the
  datastore to perform them on state on its behalf \JK{(developed contemporarily with ~\cite{S6})}. Developers can
  also load custom operations. The benefit of this approach is that NF
  instances do not have to contend for shared state. The datastore
  serializes operations issued by different instances for the same
  shared state object and applies them in the background (\secref{secproof:crossflow} proves that updates will always result in  consistent state.). This offers
  vastly better performance than the natural approach of acquiring a
  lock on state, reading it, updating, writing it back, and
  releasing the lock (\secref{s:eval}).
\begin{table}
\centering
\footnotesize
\setlength{\tabcolsep}{0.5em}
\begin{tabular}{p{2.4cm}|p{4.6cm}}

\textbf{Operation} & \textbf{Description} \\
\hline\hline
Increment/ decrement a value &  Increment or decrement the value stored at key by the given value.    \\ 
\hline
Push/pop a value to/from list & Push or pop the value in/from the list stored at the given key. \\
\hline
Compare and update &  Update the value, if the condition is true.\\

\end{tabular}
\compactcaption{Basic operations offloaded to datastore manager}
\label{t:operations}
\vspace{-0.1in}
\end{table}




%

\label{IDS-example}


\label{g:rexmit}

{\bf Non-blocking updates:} In many cases, upon receiving a packet, an
NF updates state, but does not use (read) the updated value; e.g.,
typical packet counters (e.g.,~\cite{broids,prads,squid}) are updated
every input packet, but the updated value is only read infrequently.
For such state that is written mostly and read rarely, we offer {\em
  non-blocking updates} (\tabref{ts:iostrategies}): the datastore
immediately sends the requesting instance an ACK for the operation,
and applies the update in the background.  As a further optimization,
NFs do not even wait for the ACK of a non-blocking operation; the
framework handles operation retransmission if an ACK is not received
before a timeout. \JK{If an instance wishes to read a value, the
  datastore applies all previous outstanding updates to the value, in
  the order NFs issued them, before serving the read.}

\noindent

{\bf Caching:} For all the objects which are not amenable to
non-blocking updates, we improve state access performance using novel
caching strategies that leverage state objects' scope and access
patterns (ready-heavy vs. not).
\label{caching}

{\em Per-flow state:} \name's scope-aware partitioning ensures that flows that update 
per-flow state objects are processed by a single instance; thus, these objects do not have 
cross-instance consistency requirements. The datastore's client-side library caches them at the relevant 
instance, which improves state update latency and throughput. However, for fault tolerance, we require 
local updates made to cached objects to be flushed to the store; to improve performance, these 
flush operations have non-blocking semantics (\tabref{ts:iostrategies}). 


{\em Cross-flow state:} Cross-flow state objects can be updated by
multiple instances simultaneously. Unlike prior works that
largely ignore such state, \name supports high performance shared
state management.  \CR{Some shared objects are rarely updated;
  developers can identify such objects as read-heavy. \name (1) caches such an object at the instances needing them; and (2) the client-side library at each of these instances registers a {\em callback} with the store, which is invoked whenever the store updates the object on behalf of another instance. The NF developer does not need to handle callback to update state; they are handled by the client-side library.}
  
 The cached objects only {\em serve read requests}. Whenever an (rare) update is issued by an instance - operation is immediately sent to the store, The store applies the operation and sends back the updated object to the update initiator. At the same time, the client-side library of other instances receives callback from the store and updates the locally cached value (\tabref{ts:iostrategies}). We prove this approach results in consistent updates to shared state in
 \secref{secproof:cachedcrossflow}.

%


For other cross-flow objects (not rarely-updated), the datastore
allows them to be cached at an instance only as long as no other
instance is accessing them (\tabref{ts:iostrategies}); otherwise, the objects
are flushed. \name{} notifies the client-side library when to
cache or flush the state based on (changes to) the traffic
partitioning at the immediate upstream splitter.

For {\bf scale and fault tolerance} we use multiple datastore
instances, each handling state for a subset of NF instances. \JK{Each datastore instance is multi-threaded. A thread can handle multiple state objects; however, each state object is only handled by a single thread to avoid locking overhead. }

	\section{Correctness}
\label{design}
\label{runtime-management}

So far, we focused on state management and its performance. We also
showed how \name supports requirement R1 (state availability) by
design.  We now show how it supports the requirements R2--R6. This is
made challenging both by shared state and by chaining. To support
R2-R6, \name maintains/adds metadata at the datastore, NFs and to
packets. We first describe how the most basic of the metadata --
logical packet clocks and packet logs -- are maintained. We describe
other metadata along with the requirements they most pertain to.

{\bf Logical clocks, logging:} The root (\secref{scope-aware})
attaches with every input packet a unique {\em logical clock} that is
incremented per packet.  The root also {\em logs} in the datastore
each packet, the packet clock, and to which immediate downstream
instance the packet was forwarded.  When the last NF in a chain is
done processing a packet, updating state and generating relevant
output, it informs the \name framework. \name sends a ``delete''
request with the packet's clock to the root which then removes the
packet from the log. Thus, at any time, the root logs all packets that
are being processed by one or more chain instances. \JK{When any NF in the
chain cannot handle the traffic rate, the root log builds in size; CHC
drops packets at the root when this size crosses a threshold to avoid
buffer bloat.} When multiple root instances are in use
(\secref{scope-aware}), we encode the identifier of the root instance
into the higher order bits of the logical clock inserted by it to help
the framework deliver ``delete'' requests to the appropriate root
instance.

\subsection{R2, R3: Elastic scaling}
\label{g:scaling}

In some situations, we may need to reallocate ongoing processing of
traffic across instances. This arises, e.g., in elastic scaling, where
a flow may be processed at an ``old'' instance and reallocated to a
``new'' scaled up instance.  We must ensure here that the old and new
instances operate on the correct values of per- and cross-flow state
even as traffic is reassigned (requirements R2 and R3). 

Specifically, for cross-flow shared state, we require that: {\em Updates made to the shared state by every incoming packet are
  reflected in a globally consistent order irrespective of which NF
  instance processed the corresponding packet.}

Existing systems achieve this at high overhead: OpenNF~\cite{opennf}
copies shared internal state from/to the instances sharing it, each
time it is updated by an incoming packet! In contrast, ensuring this
property in \name is straightforward due to externalization and
operation offloading (\secref{storage-sec}): when multiple instances
issue update operations for shared state, the datastore serializes
the operations and applies in the background. All subsequent accesses to the
shared state then read a consistent state value.

Per-flow state's handling must be correctly reallocated across
instances, too (R2).  One approach is to disassociate
the old instance from the state object (by having the instance remove
its instance{\_}ID from the object's metadata) and associate the new
instance (by adding its instance{\_}ID).  But, this does not ensure
correct handover when there are in-transit packets that update the
state: even if the upstream splitter immediately updates the
partitioning rules and the traffic starts reaching the new instance,
there might be packets in-transit to, or buffered within, the old
instance. If the new instance starts processing incoming packets right
away then state updates due to in-flight/buffered packets may be disallowed
by the datastore (as a new instance is now associated with the state
object) and hence the updates will be lost.

Thus, to satisfy R2, we require: {\em Loss-freeness, i.e., the state
  update due to every incoming packet must be reflected in the state
  object.} Furthermore, some NFs may also need {\em
  order-preservation: updates must happen in the order of packet
arrivals into the network}.

  These properties are crucial for off-path NFs, e.g., IDS. Such NFs
  cannot rely on end-to-end retransmissions to recover from lost
  updates induced by traffic reallocation~\cite{opennf}. Similarly,
  they may have to process packets in the order in which they are
  exchanged across two directions of a flow, and may be thwarted
  by a reordering induced by reallocation (resulting in false
  positives/negatives).

\begin{figure}[t!]
	\begin{center}
		\includegraphics[width=0.45\textwidth]{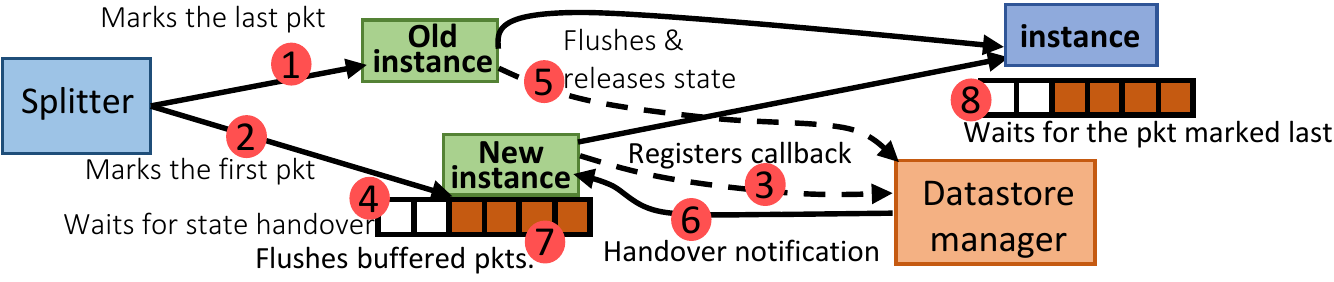}
		\compactcaption{State handover.}
		\label{fig-scaling}
	\end{center}   
	\vspace{-0.2in}     
\end{figure}

\figref{fig-scaling} shows the sequence of steps \name takes for R2:
\circled{1} The splitter marks the ``last'' packet sent to the old
instance to inform the old instance that the flow has been moved. This
mark indicates to the old instance that it should flush any cached
state associated with the particular flow(s) to the datastore and
disassociate its ID from the per flow state, once it has processed the ``last'' packet.  \circled{2} The splitter
also marks the ``first'' packet from the traffic being moved
to the new instance. \circled{3} When the new instance receives the
``first'' packet, it tries to access the per flow state from the
datastore. If the state is still associated with the old
instance{\_}ID, it registers a callback with the datastore to be
notified of metadata updates. \circled{4} The new instance starts
buffering all the packets associated with the flow which is being
moved.  \circled{5} After processing the packet marked as ``last'',
the old instance flushes the cached state and updates the metadata to
disassociate itself from the state. \circled{6} The datastore notifies
the new instance about the state handover.  \circled{7} The new
instance associates its ID with the state, and flushes its buffered
packets.

The above ensure that updates are not lost {\em and} that they
happen in the order in which packets arrived at the
upstream splitter. In contrast, OpenNF provides {\em separate
  algorithms} for loss-freeness and order-preservation; an NF author
has the arduous task of choosing from them!

Note also that packets may arrive out of order at a {\em downstream}
instance, causing it to make out-of-order state updates. To prevent
this: \circled{8} The framework ensures that packets of the moved flow emitted
by the new instance are not enqueued at the downstream instance, but
instead are buffered internally within the framework until the 
packet marked as ``last'' from the old instance is enqueued at
the new instance.

\subsection{R4: Chain-wide ordering}
\label{add}
\label{trojan_example}
\label{chainwide-ordering}

To support R4, we require that:
{\em Any NF in a chain should be able to process packets, potentially spread
  across flows, in the order in which they entered the NF chain.}
\name's logical clocks naturally allow NFs to reason about
cross-flow chain-wide ordering  and satisfy R4. E.g., the
Trojan detector from \secref{requirements} can use packets' logical clocks to
determine the arrival order of SSH, FTP and IRC connections.

\subsection{R5: Straggler mitigation}
\label{straggler}
\label{stragglers}
R5 calls for the following:
{\em All duplicate outputs, duplicate state updates, and duplicate
  processing are suppressed.}

A key scenario in which duplicate suppression is needed is straggler
mitigation.  A straggler is a slow NF that causes the entire NF
chain's performance to suffer. We first describe \name's mechanism for
straggler mitigation (which kicks in once user-provided logic
identifies stragglers;~\secref{sec:api}), followed by duplicate
suppression.

{\bf Clone and replay:} To mitigate stragglers \name deploys {\em
  clones}. A clone instance processes the same input as the original
in parallel. \name retains the faster instance, killing the
other. \name{} initializes the clone with the straggler's latest
state from the datastore. It then replicates incoming traffic from the
upstream splitter to the straggler and the clone.

This in itself is not enough, because we need to satisfy R2, i.e.,
ensure that the state updates due to packets that were in-transit to
the straggler at the time the clone's state was initialized are
reflected in the state that the clone accesses. To address this, we
{\em replay} all logged packets from the root. The root continues to
forward  new incoming packets alongside replayed ones. The clone
processes replayed traffic first, and the framework buffers replicated
traffic. To indicate end of replay traffic, the root marks the
``last'' replayed packet (this is the most recent logged packet at the
time the root started replaying).  When replay ends (i.e., the packet
marked ``last'' was processed by the clone), the framework hands buffered
packets to the clone for processing.

Given the above approach for straggler mitigation, there are three
forms of duplicates that can arise. \name suppresses them by
maintaining suitable metadata.

{\bf 1. Duplicate outputs:} Replicating input to the clone results in
duplicate outputs. Here, the framework suppresses duplicate
outputs associated with the same logical clock at message queue(s)
of immediate downstream instance(s).

\begin{figure}[]
	\centering
	\subfloat[][Naive reprocessing]{%
		\label{fig-ft-nv-reproc}%
		\includegraphics[scale=0.57]{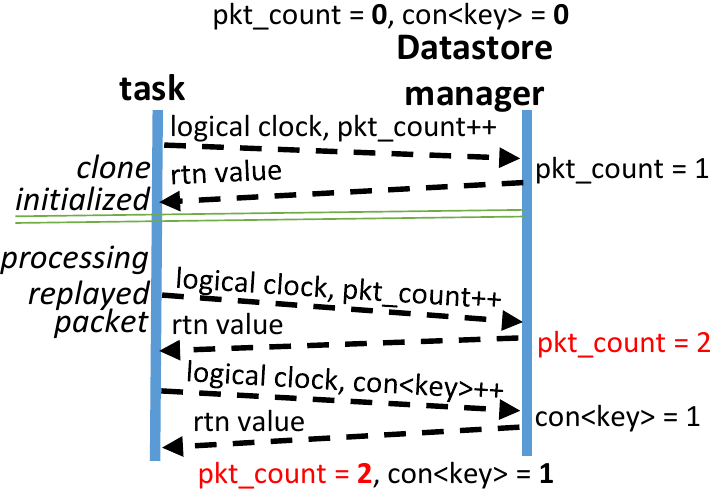}%
	}
	\hspace{0em}
	\subfloat[][Reprocessing with emulation]{%
		\label{fig-ft-reproc} 
		\includegraphics[scale=0.57]{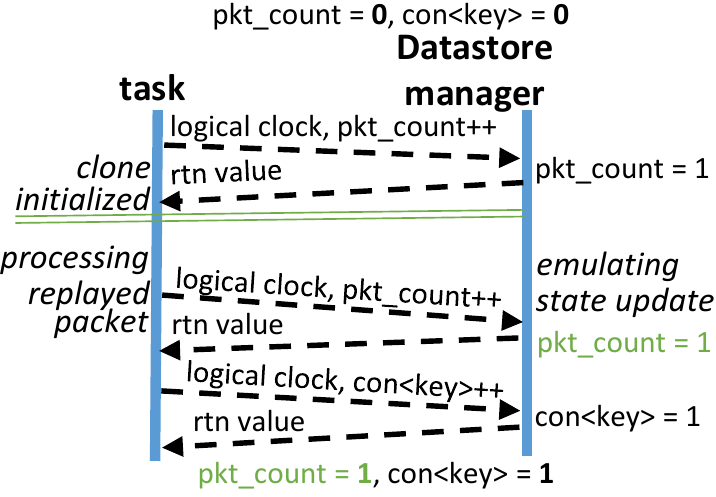}%
	}

	\compactcaption{Duplicate update suppression}
	\label{fig-ft}
\end{figure}

{\bf 2. Duplicate state updates:} Some of the replayed packets may have already 
updated some of the stragglers' state objects.  For example, an
IDS updates both the total packet count and
the number of active connections per host. A clone IDS may have been
initialized after the straggler updated the former but not the latter.
In such cases, processing a replayed packet can {\em incorrectly}
update the same state (total packet count) multiple times at the straggler (\figref{fig-ft-nv-reproc}).
To address this, the datastore logs the state value
corresponding to each state update request issued by any instance, as well
as the logical clock of the corresponding packet. This is only done
for packets that are currently being processed by some NF in the chain.
During replay, when the straggler or clone sends an update for a state object, the
datastore checks if an update corresponding to the logical
clock of the replayed packet has already been applied; if so, the
datastore {\em emulates} the execution of the update by
returning the value corresponding to the update
(\figref{fig-ft-reproc}). \JK{In  Appendix \ref{non-deter}, we describe how \name handles non-deterministic state update operations.}





{\bf 3. Duplicate upstream processing:} NFs upstream from the
clone/straggler would have already processed some of the in-transit
packets. In such cases, reprocessing replayed packets leads to
incorrect actions at upstream NFs (e.g., an IDS may raise false
alarms). 
To address this, each replayed packet is marked and it
carries the ID of the clone where it will be processed. Such
packets need special handling: the intervening instances
recognize that they are not suspicious duplicates; if
necessary, the instances read the store for state corresponding
to the replayed packet, make any needed modifications to the
packet's headers, and produce relevant output;
the instances can issue updates to state, too, but in such
cases the datastore {\em emulates} updates as before.
The clone's ID is cleared once it processed the packet.

\subsection{R6: Safe Fault Recovery}
\label{faulttolerance}
\label{s:ft-in}


\JK{ Our description of R6 in \secref{s:mot} focused on NF failures;
  however, since \name introduces framework components, we generalize
  R6 to cover other failures as well. Specifically, we require the
  following general guarantee:

{\bf Safe recovery Guarantee:} {\em When an NF instance or a framework component fails and a recovery occurs, we must ensure that the state at each NF in the chain has the same value as under no failure.}

We assume the standard fail-stop model,  that a machine/node can crash at any time and that the other machines/nodes in the system can immediately detect the failure.

First, we show how \name leverages metadata to handle the failure of individual components. Then, we discuss scenarios involving simultaneous failure of multiple components.
}
  
{\bf NF Failover:} When an NF fails, a failover instance takes over
the failed instance's processing. The datastore manager associates the
failover instance's ID with relevant state. Packet replay brings state
up-to-speed (from updates due to in-transit packets). Similar to
cloning (\secref{stragglers}), we suppress duplicate state
updates and upstream processing.

Since ``delete'' requests are generated after the last NF is done
processing a packet, failure of such an NF needs special handling:
consider such an instance T failing after generating an output packet
for some input packet P, but before the framework sends a ``delete''
request for P.  When P is replayed, T's failover instance produces
output again, resulting in duplicate packets at the receiving end
host. To prevent this, for the last NF in the chain, our framework
sends the ``delete'' request for P {\em before} the NF generates the
output packet. If the NF fails before the ``delete'' request, then P
will be replayed, but this does not result in duplicate downstream
processing since the NF did not generate output.  If the NF fails
after the ``delete'' request but before generating output, then P is
not replayed, and hence the end host will not receive any output
packet corresponding to P. \CR{To the host, this will appear as a packet being dropped by the network, causing P to be retransmitted from the source and resulting in
correct overall behavior.} In \secref{secproof:nfrecovery}, we show that using this
  protocol an NF instance recovers with state similar to that under
  no failure.

\begin{figure}[tb]
	\vspace{-.1in}
	\centering
		\includegraphics[scale=0.68]{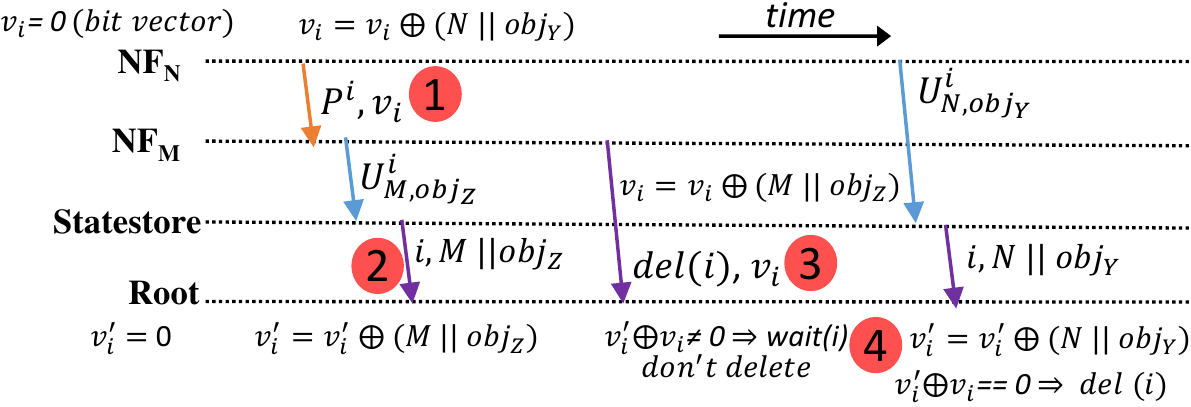}%
                \compactcaption{Recovery under
                non-blocking operations. Consider a packet $P^i$ which
                is processed by $NF_N$, followed by $NF_M$, the last NF in the
                chain. $NF_N$ and $NF_M$ update objects $obj_Y$ and $obj_Z$, respectively.}
                
	\label{fig-nonblocking-ft}
	\vspace{-0.08in}
\end{figure}

{\bf Non-blocking operations:} Non-blocking updates, where NF
instances don't wait for ACKs, instead relying on the framework to
handle reliable delivery, can introduce the following failure mode: a
instance may fail after issuing state update but before the
update is committed and an ACK was received. In such cases, to ensure
R6, we need that {\em the framework must re-execute the incomplete
  update operation.}

Suppose an instance $N$ fails after processing packet $P_i$ ($i$ is the
logical clock) but before the corresponding state update operation
$U_{N,obj}^i$ ($obj$ is the state object ID) completes. $P_i$ may have
induced such operations at a subset of NF instances $\{N\}$ along the
chain.  A natural idea to ensure the above property is to replay
packets from the root to reproduce $U_{N,obj}^i$ at various $N$'s. For
this, however, $P_i$ must be logged and should not have been
deleted. If $P_i$ is deleted it can't be replayed.

We need to ensure $P_i$ continues to be logged as long as there is
some $N$ for which $U_{N,obj}^i$ is not committed. Our approach for
this is shown in~\figref{fig-nonblocking-ft}: \circled{1} Each packet carries
a 32-bit vector $v_i$ (object ID and instance ID; 16b each) that is
initialized to zero. Each NF instance where processing the packet
resulted in a state update XORs the concatenation of its ID and the
corresponding state objects' IDs into the bit vector.  \circled{2} When
committing a given NF's state update, the state store signals to the
root the clock value of the packet that induced the update as
well as the concatenated IDs. \circled{3} The last instance sends the final
vector along with its ``delete'' request to the root. \circled{4} When a delete
request and the final vector are received, the root XORs the
concatenated IDs with the concatenated IDs reported by each signal
from the state store in step 2. If the result is zero, this implies
that updates induced by the packet at all NF instances $\{N\}$ were
committed to the store; the root then proceeds to delete the
packet from the log. Otherwise, the packet updated state at some NF,
but the NF has not yet reported that the state was committed; here,
the root does not delete the packet.


{\bf Root:} To ensure R6 under root failover, we need that a new root
must start with the logical clock value and current flow allocation at
the time of root failure. This is so that the new root processes
subsequent packets correctly. To ensure this, the failover root reads
the last updated value of the logical clock from the datastore, and
retrieves how to partition traffic by querying downstream instances'
flow allocation. The framework buffers incoming packets during root
recovery. \JK{We prove this approach ensures recovery with a state similar to that under no failure in   \secref{secproof:rootrecovery}.}

{\bf Datastore instance:} Recall that different NFs can store their
states in different storage instances (\secref{storage-sec}). This
ensures that store failures impact availability of only a portion of
the overall state being tracked. Now, to ensure R6 under the failure
of a datastore instance, we need that the recovered state in the new
store instance must represent the value which would have resulted if
there was no failure. The recovered state must also be consistent with
the NF instances' view of packet processing thus far (i.e., until
failure).

To support this property we distinguish between per-flow and shared
state. For the former, we leverage the insight that all the NFs
already maintain an updated cached copy of per-flow state. If a
datastore instance fails, we can simply query the last updated value of the
cached per-flow state from all NF instances that were using the store.


\JK{Recovering shared state is nuanced. For this, we use
checkpointing with write-ahead logging~\cite{wal}. The datastore periodically checkpoints
shared state along with the metadata, ``$TS$'', which is the set of
logical clocks of the packets corresponding to the last state
operation executed by the store on behalf of each NF instance. Each
instance locally writes shared-state update operations in a write-ahead log}. Say the latest checkpoint was at time $t$ and failure
happens at $t + \delta$. A failover datastore instance boots with
state from the checkpoint at $t$. This state now needs to be ``rolled
forward'' to $t+\delta$ and made consistent with the NF instances'
view of packet processing at $t+\delta$. Two cases arise:

(Case 1) If NF instances that were using the store instance don't read
shared state in the $\delta$ time interval, then to recover shared
state, the framework re-executes state update operations from the local write-ahead log on behalf of each NF, starting from the logical clocks
included in the metadata $TS$ in the checkpoint. Recall that in our
design the store applies updates in the background, and this update
order is unknown to NF instances. Thus, our approach ensures that the
state updates upon re-execution match that produced by a {\em
  plausible} sequence of updates that the store may have enforced
prior to failure. This consistency property suffices because, in
Case 1, NFs are not reading shared state in the $\delta$ interval.

\begin{figure}[]
	\begin{center}
		\includegraphics[width=0.42\textwidth]{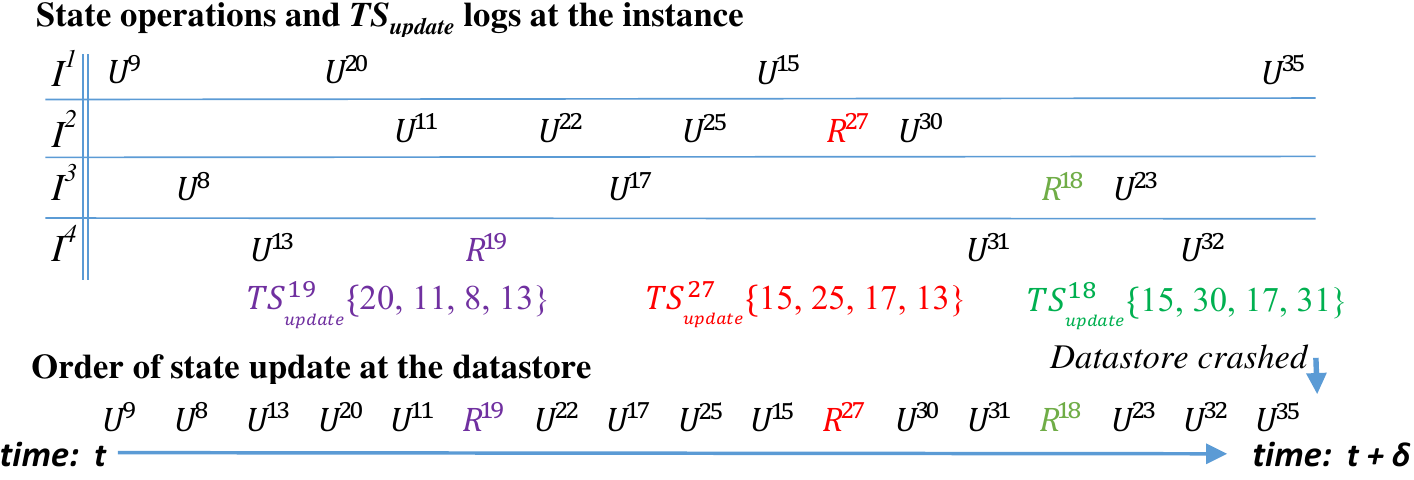}
                \compactcaption{Recovering shared state at the
                  datastore. $I^k$ are instances.
                  $U^{logical\_clock}$ and $R^{logical\_clock}$
                  represent ``update'' and ``read''.}
		\label{fig-dsfailure}
	\end{center}  
	    \vspace{-1.2em}      
\end{figure}

(Case 2) Say an NF instance issues a read between $t$ and $t +
\delta$; e.g., $I^3$ in \figref{fig-dsfailure} issues
$R^{18}$. Following the above approach may lead to an order of
re-execution such that the actual state $I^3$ read in $R^{18}$ is
different from the state in the store after recovery. \JK{To ensure that
the store's state is consistent with all $I^k$'s current view, the
framework must re-execute operations in such an order that the
datastore would have produced the same value for each read in $[t,
  t+\delta]$.}

To ensure this, on every read operation, the datastore returns $TS$
along with the latest value of the shared state (e.g., $TS^{19}$ is
returned with $I^4$'s $R^{19}$). The instance then logs the value of
the shared state along with the corresponding $TS$.  Re-execution upon
failure then needs to select, among all $TS$'s at different instances,
the one corresponding to the most recent read from the store prior to
the crash (i.e., $TS^{18}$, since $R^{18}$ in the most recent read;
most recent clock does not correspond to most recent read). How
selection is done is explained shortly; but note that when the
framework re-executes updates starting from the clock values indicated
by this selected $TS$ that would bring the store in sync with all
NFs. In our example, $TS^{18}$ is the selected $TS$; we
initialize the store state with the value in the corresponding read
($R^{18}$).  From the write-ahead log of each NF, the framework re-executes
update operations that come after their corresponding logical clocks
in $TS^{18}$. At instance $I^1$, this is the update after $U^{15}$,
i.e., $U^{35}$. At  $I^3$ and $I^4$ these are $U^{23}$ and
$U^{32}$, respectively. Shared state is now in sync with all NFs.


$TS$ selection works as follows: first we form a set of all the $TS$'s
at each instance, i.e., $Set = \{TS^{18},TS^{19},TS^{27}\}$. Since the
log of operations at an instance follows a strict clock order we
traverse it in the reverse order to find the latest update operation
whose corresponding logical clock value is in $Set$.  For example, if
we traverse the log of $I^1$, we find that the logical clock of
$U^{15}$ exists in $Set$. After identifying such a logical clock
value, we remove all the entries from $Set$ which do not contain the
particular logical clock value (such $TS$s cannot have the most recent
read); e.g., we remove $TS^{19}$ as it does not contain logical clock
$15$. Similarly, we remove $TS^{27}$, after traversing $I^2$'s
log. Upon doing this for all instances we end up selecting $TS^{18}$
for recovery. In \secref{secproof:storerecovery}, we prove that using this
protocol the store recovers with state similar to that under no failure.




{\bf Correlated failures:} Using the above approaches, \name can also
handle correlated failures (\tabref{ts:mutliplefailures}) of multiple
NF instances, root, and storage instances. However, \name
cannot withstand correlated failure of a store instance with any other
component that has stored its state in that particular
instance. Replication of store instances can help recover from such
correlated failures, but that comes at the cost of increasing the per
packet processing latency.

\begin{table}
	\centering
	\setlength{\tabcolsep}{0.5em}
        \small
		\begin{tabular}{l||cc}
			&
			NF instance&
			Root\\
			
			\hline
			\hline
			
			Store instance & \cmark$^*$ & \cmark$^*$\\
			NF instance & \cmark &\cmark\\
			
		\end{tabular}
		\caption{Handling of correlated failures ($^*$Cannot recover if component and the store instance storing its state fail together).}
	\vspace{-1.2em}
\label{ts:mutliplefailures}	
\end{table}






	\vspace*{-0.1in}
\section{Implementation}
Our prototype consists of an execution framework and a datastore,
implemented in C++. NFs runs in LXC containers~\cite{lxc} as 
multithreaded processes. NFs are implemented using our \name library
that provides support for input message queues, client side datastore
handling, retransmissions of un-ACK'd state updates
(\secref{g:rexmit}), statistics monitoring and state handling. Packet
reception, transmission, processing and datastore connection
are handled by different threads.

For low latency, we leverage Mellanox messaging accelerator
(VMA)~\cite{vma} which enables user-space networking with kernel
bypass similar to DPDK~\cite{dpdk}. In addition to this, VMA also
supports TCP/UDP/IP networking protocols and does not require any
application modification. Even though we use VMA, we expect similar
performance with other standard kernel bypass techniques.
Protobuf-c~\cite{protobuf} is used to encode and decode messages
between a NF instance and the datastore. Each NF instance is
configured to connect to a ``framework manager'' to receive information
about it's downstream instances (to which it connects via tunnels),
datastore instances and other control information.

The framework manager can dynamically change the NF chain by
instantiating new types of NFs or NF instances and updating
partitioning information in upstream splitters\footnote{based on statistics from vertex managers}. Our datastore
implements an in-memory key-value store and supports the operations
in~\tabref{t:operations}. We reimplemented four NFs atop
\name. Table~\ref{ts:nf-state} shows their state objects, along with
the state's scope and access patterns.

\begin{table}[]
	\centering
    \scriptsize
	\begin{tabular}{p{0.8cm}||p{2.9cm}|p{3.65cm}}

		{\bf NF} & {\bf Description of state object} & {\bf Scope; access pattern} \\ 
		\hline 
		\hline
		\multirow{2}{*}{}
		& Available ports& Cross-flow; write/read often \\ \cline{2-3} 
		NAT& Total TCP packets & Cross-flow; write mostly, read rarely \\ \cline{2-3} 
		& Total packets& Cross-flow; write mostly, read rarely \\ \cline{2-3} 
		& Per conn. port mapping&  Per-flow; write rarely, read mostly\\ 
		\hline 
		\hline
 		\multirow{2}{*}{}Trojan detector& Arrival time of IRC, FTP and SSH flows for each host   &Cross-flow;
 		write/read often\\ 
 		\hline 
 		\hline
 		\multirow{2}{*}{}Portscan detector&Likelihood of being malicious (per host)&Cross-flow; write/read often\\ \cline{2-3} 
 		& Pending conn. initiation req. along with its timestamp & Per-flow; write/read often \\ 
 		\hline 
 		\hline
 		\multirow{3}{*}{}Load& Per server active \# of conn.&Cross-flow; write/read often \\ \cline{2-3} 
 		balancer&  Per server byte counter&Cross-flow; write mostly, read rarely\\ \cline{2-3} 
 		& Conn. to server mapping & Per-flow; write rarely, read mostly \\ 
	\end{tabular}
		\compactcaption{NFs and description of their state objects	}
		\label{ts:nf-state}
\vspace{-0.1in}
\end{table}

{\bf NAT:} maintains the dynamic list of available ports in
the datastore. \JK{When a new connection arrives, it obtains an available
port from the datastore (The datastore pops an entry from the list of available ports on behalf of the NF)}. It then updates: 1) per-connection
port mapping (only once) and, 2)  (every packet) L3/L4 packet counters.

{\bf Portscan detector}~\cite{trw}:  detects
infected port scanner hosts. It tracks new connection
initiation for each host and whether it was successful or not. On each
 connection attempt, it updates the
likelihood of a host being malicious, and blocks a host when the
likelihood crosses a threshold.


{\bf Trojan detector:} implementation here follows~\cite{lorenzos_work}.

{\bf Load balancer:} maintains the load on each
backend server. Upon a new connection, it
obtains the IP of the least loaded server from the datastore and
increments load. It then updates: 1) connection-to-server mapping
2) per server \#connections and, 3) (every packet) per server byte counter
.



	\vspace{-0.1in}
\section{Evaluation}
\label{eval}
\label{s:eval}

 We use two packet traces (Trace\{1,2\}) collected on the link between
 our institution and AWS EC2 for  trace-driven evaluation of our
 prototype. Trace1 has 3.8M packets with 1.7K connections and Trace2
 has 6.4M packets with 199K connections. The median packet sizes are
 368B and 1434B. We conducted all experiments with both traces and
 found the results to be similar; for brevity, we only show results
 from Trace2. We use six CloudLab~\cite{cloudlab} servers each with
 8-core Intel Xeon-D1548 CPUs and a dual-port 10G NIC. One port is
 used to forward traffic, and the other for datastore communication
 and control messages. To process at 10Gbps, each NF instance runs
 multiple processing threads. \name performs scope-aware partitioning
 of input traffic between these threads. Our datastore runs on a
 dedicated server.



\vspace{-.1in}
\subsection{State Management Performance}
\label{extern_state}

{\bf Externalization:} We study three models which reflect the state
access optimizations discussed in (\secref{storage-sec}): \#1) All
state is externalized and non-blocking operations are used. \#2)
Further, NFs cache relevant state objects. \#3) Further, NFs do not
wait for ACKs of non-blocking operations to state objects; the
framework is responsible for retransmission
(\secref{storage-sec}). \CR{The state objects per NF that benefit from \#2
and \#3 can be inferred from \tabref{ts:iostrategies}	and \tabref{ts:nf-state}; e.g., for NAT,
per-connection port mapping is cached in \#2, and
the two packet counters benefit from non-blocking updates in \#3. We
compare these models with a ``traditional'' NF where all state is
NF-local. We study each NF type in isolation first.}

\begin{figure}[t!]
	\begin{center}
		\includegraphics[width=0.45\textwidth]{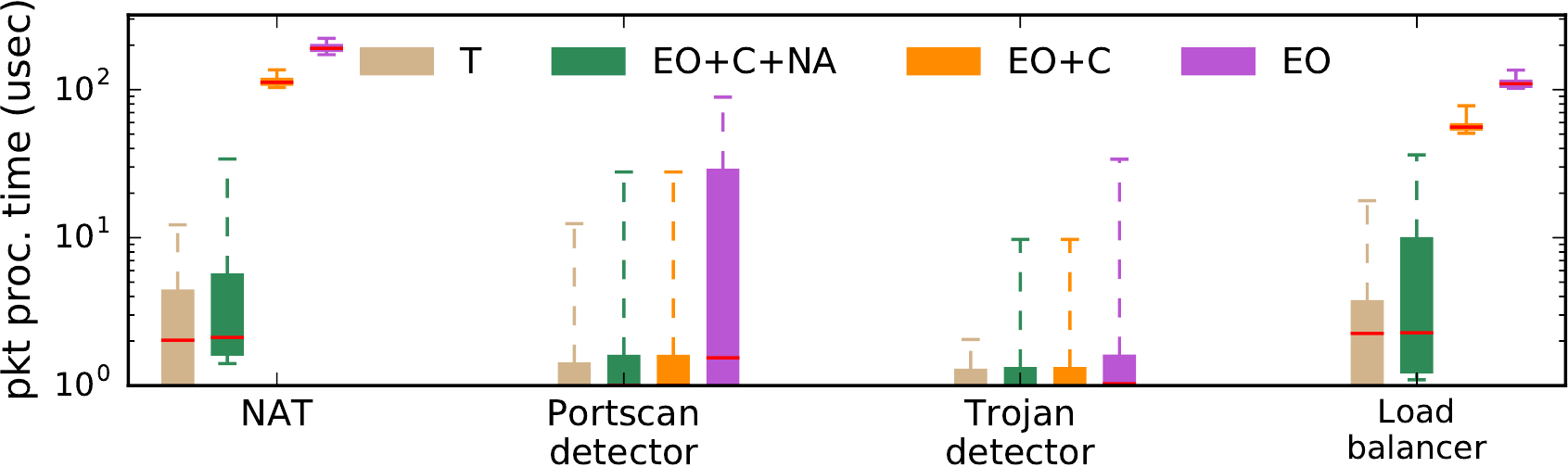}
		\\
			\vspace{-0.1em}

	\compactcaption{5\%ile, 25\%ile, median, 75\%ile and 95\%ile pkt processing times. ({\footnotesize T = Traditional NF, EO = Externalized state operations, C = with caching, NA = without waiting for the ACK}) }
	\label{eval-micro-state-exten}
	\end{center}      
		\vspace{-0.2in}  
\end{figure}

%

~\figref{eval-micro-state-exten} shows the per packet
processing times.
The median times for traditional NAT and load balancer are 2.07$\mu$s
and 2.25$\mu$s, respectively. In model \#1, this increases by 190.67$\mu$s
and 109.87$\mu$s, respectively, with network RTT contributing to most of
this (e.g., NAT needs three RTTs on average per packet: one for
reading the port mapping and other two for updating the two
counters). We don't see a noticeable impact for scan and Trojan
detectors (they don't update state on every packet).

Relative to \#1, caching (\#2) further lowers median processing times
by 111.98$\mu$s and 55.94$\mu$s for NAT\footnote{NAT needs 2 RTTs to
  update counters as port mapping is cached.} and load
balancer. For portscan and Trojan detector, reduces it by 0.54$\mu$s and 0.1$\mu$s (overhead becomes {\bf+0.1$\mu$s} as compared to traditional NFs) as \name caches the cross-flow state. Later, we evaluate the benefits of cross-flow caching in detail. Finally, \#3 results in median packet processing times of {\bf 2.61$\mu$s} for NAT (which now needs 0 RTTs on average)
and {\bf 2.27$\mu$s} for load balancer. These represent small
overheads compared to traditional NFs: {\bf +0.54$\mu$s} for NAT, and
{\bf +0.02$\mu$s} for the load balancer (at the median). Note that for portscan and Trojan detector the performance of \#3 is comparable to \#2 as they don't have any blocking operations.

We constructed a simple chain consisting of one instance each of NAT,
portscan detector and load balancer in sequence, and the Trojan
detector operating off-path attached to the NAT. With model \#3, the
median end-to-end overhead was {\bf 11.3$\mu$sec} compared to using
traditional NFs.



		

{\bf Operation offloading:} We compare \name's operation offloading
against a naive approach where an NF first reads state from
the datastore, updates it, and then writes it back. We turn off
caching optimizations. We now use two NAT instances updating shared state
(available ports and counters). We find that the median packet
processing latency of the naive approach is {\bf 2.17X} worse
(64.6$\mu$s vs 29.7$\mu$s), because it not only requires 2 RTTs to
update state (one for reading the state and the other for writing it
back), but it may also have NFs wait to acquire locks.
\name's aggregate
throughput across the two instances is $>$2X better.

\begin{figure}[t!]
	\begin{center}
		\includegraphics[width=0.47\textwidth]{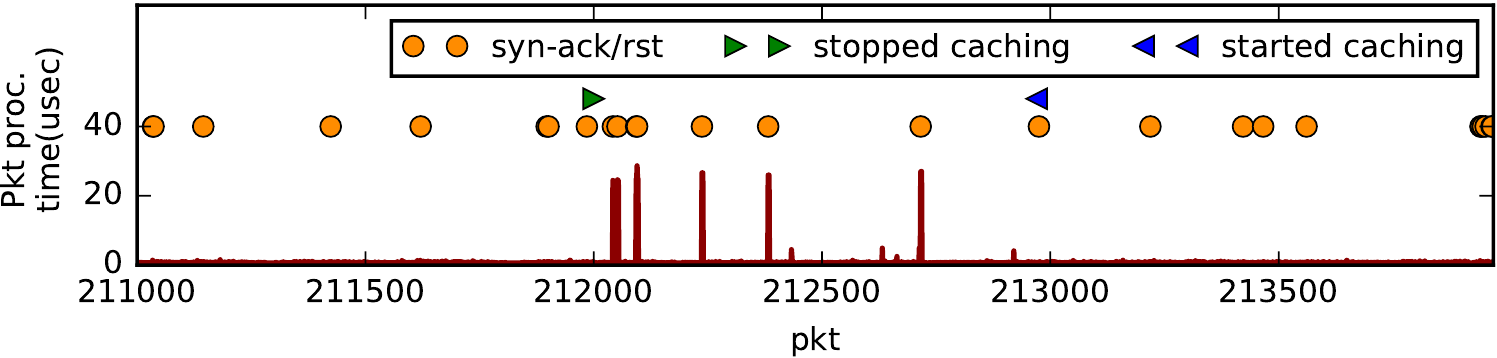}
		\\
		\compactcaption{Per packet processing latency with cross-flow state caching}
		\label{eval-crossflow}
	\end{center}      
		\vspace{-1em}  
\end{figure}

{\bf Cross-flow state caching:} To show the performance of our
cross-flow state caching schemes (\tabref{ts:iostrategies}; Col 5), we
run the following experiment: we start with a single portscan
detector. After it has processed around 212K packets, we add a second
instance and split traffic such that for particular hosts, captured by
the set $\mathcal{H}$, processing happens at both instances. At around
213K packets, we revert to using a single instance for all
processing. \figref{eval-crossflow} shows the benefits of
caching the shared state. At 212K packets, when the second instance is
added, the upstream splitter signals the original instance to flush
shared state corresponding to hosts $\in \mathcal{H}$
(\tabref{ts:nf-state}). From this point on, both instances make
blocking state update operations to update the likelihood of hosts
$\in \mathcal{H}$ being malicious on every successful/unsuccessful
connection initiation. Thus, we see an increase in per packet
processing latency for every SYN-ACK/RST packet. At packet number
213K, all processing for $\mathcal{H}$ happens at a single instance
which can start caching state again. Thus, the processing latency for
SYN-ACK/RST packets drops again, because now state update
operations are applied locally and updates are flushed in a
non-blocking fashion to the store.

\begin{figure}[t!]
	\begin{center}
		\includegraphics[width=0.45\textwidth]{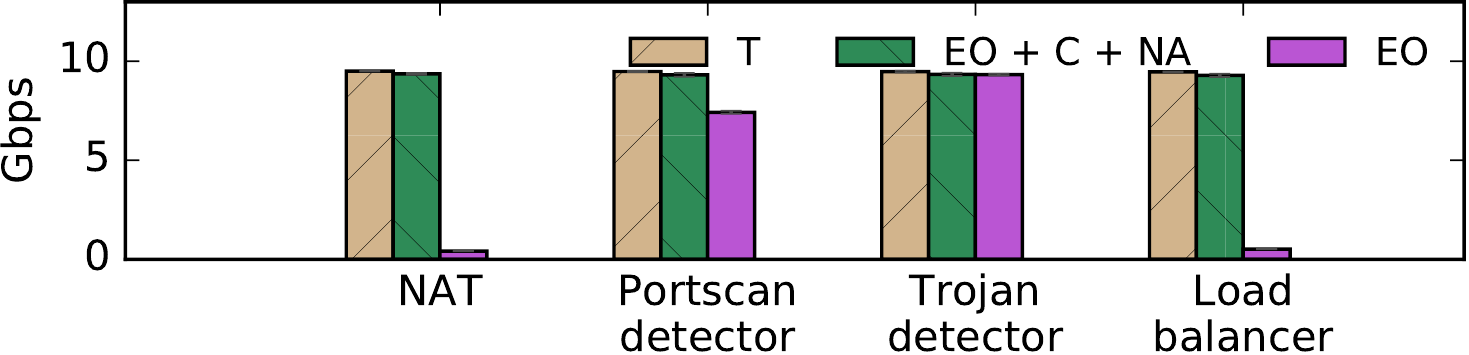}
		\\
		\compactcaption{Per instance throughput. (T = Traditional NF, EO = Externalized state operations, C = with caching, NA = without waiting for the ACK)}
		\label{eval-ft-throughput}
	\end{center}      
	\vspace{-1em}  
\end{figure}

{\bf Throughput:} We measure degradation in per NF throughput for
models \#1 and \#3 above compared to traditional
NFs. \figref{eval-ft-throughput} shows that the max. per NF
throughput for traditional NFs is around 9.5Gbps. Under model \#1,
load balancer and NAT throughput drops to 0.5Gbps. The former needs
to update a byte counter (which takes 1 RTT) on every packet;
likewise, the NAT needs three RTTs per packet. The port scan and Trojan
detectors do not experience throughput degradation because they don't
update state on every packet. Model \#3 increases throughput to
9.43Gbps, matching traditional load balancer and NAT. We repeated our
experiment with the aforementioned single-instance NF chain and
observed similar maximal performance (9.25Gbps with both \name and
traditional NFs) in Model \#3.

{\bf Datastore performance} We benchmarked the datastore using the
workload imposed by our state operations. \CR{We used 128bits key and  64bits value to benchmark the datastore. The datastore was running four threads. Each thread handled 100k unique entries. As discussed in \secref{storage-sec}, state is not shared between these threads.} We found that a single instance of our datastore supports {\bf $\sim$5.1M} ops/s (increment
at 5.1M ops/s, get at 5.2M ops/s, set at 5.1M ops/s;
~\tabref{t:operations}). The datastore can be easily scaled to support a greater
rate of operations by simply adding multiple instances; each
state object is stored at exactly one store node and hence no cross-store node
coordination is needed.

\subsection{Metadata Overhead}

{\bf Clocks:} The root writes packet clocks to the
datastore for fault tolerance. This adds a 29$\mu$s
latency per packet (dominated by RTT). We optimize further by writing
the clock to the store after every n$^{th}$ packet.\footnote{After a
  crash, this may lead the root to assign to a packet an already
  assigned clock value. To overcome this issue, the root starts
  with {\em n + last update} so that clock values
  assigned to packets represent their arrival order.}  The average
overhead per packet reduces to 3.5$\mu$s and 0.4$\mu$s for
$n=10, 100$.
	
	
	
	\label{logging_overhead}
	{\bf Packet logging:} We evaluated two models of logging: 1)
        locally at the root, 2) in the datastore.  The former
        offers better performance, adding 1$\mu$s latency per packet,
        whereas the latter adds 34.2$\mu$s but is more fault tolerant
        (for simultaneous root and NF failures).  We also studied the
        overhead imposed by the framework logging clocks and
        operations at NFs, the datastore logging clocks and state, and
        the XOR-ing of identifiers (\secref{s:ft-in}); the 
        performance impact for our chain (latency and throughput
        overhead) was negligible ($<1$\%).
        


\JK{{\bf XOR check and delete request:} (\secref{s:ft-in}) XOR checks of bit vectors are performed asynchronously in the background and do not introduce any 
latency overhead.  However, ensuring the successful delivery of ``delete'' request to root before forwarding the packet 
introduces a median latency overhead of 7.9$\mu$sec. Asynchronous ``delete'' request operation eliminates this overhead but failure of the last NF in a chain may result in duplicate packets at the receiver end host.

}
\subsection{Correctness Requirements: R1--R6}
\label{statemanagement-eval}

\begin{figure}[!t]
	\begin{minipage}[t]{0.4\linewidth}
		\centering
		\includegraphics[width=1\textwidth]{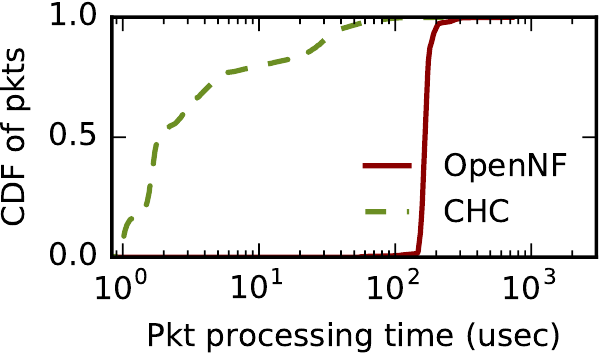}
		\caption{State sharing.}
		\label{eval-scaling}
	\end{minipage}
	\hspace{0.1cm}
	\begin{minipage}[t]{0.4\linewidth} 
		\centering
		\includegraphics[width=1\textwidth]{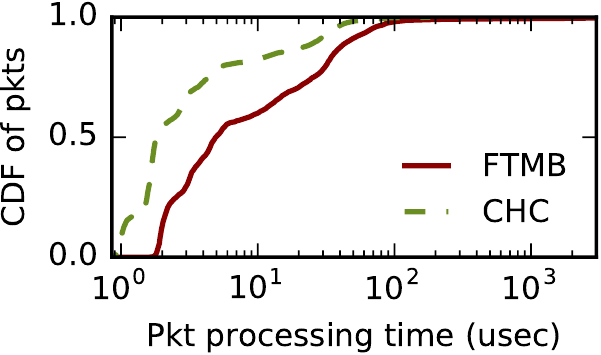}
		\caption{Fault recovery.}
		\label{eval-ft}
	\end{minipage}        
\vspace{-0.12in}
\end{figure}

{\bf R1: State availability:} Using our NAT, we compare
FTMB's~\cite{ftmb} checkpointing approach with \name writing all state
to a store.  We could not obtain access to FTMB's code; thus, we
emulate its checkpointing overhead using a queuing delay of 5000$\mu$s
after every 200ms (from Figure 6 in \cite{ftmb}).
\figref{eval-ft} (with 50\% load level) shows that checkpointing in
FTMB has a significant impact: the 75th\%-ile latency is 25.5$\mu$sec
-- which is {\bf 6X} worse than that under \name (median is 2.7X
worse).  FTMB's checkpointing causes incoming packets to be
buffered. Because of externalization in \name, there is no need for such
checkpointing.
Also, FTMB does not support recovery of the packet
logger~\cite{ftmb}. \name intrinsically supports this
(\secref{s:ft-in}), and we evaluate it in \secref{s:ft-eval}.

{\bf R2: Cross-instance state transfers:} We elastically scale up NAT as follows: We replay our trace for 30s through a single instance;
midway through replay, we reallocate 4000 flows to a new instance,
forcing a move of the state corresponding to these flows.  We compare
\name with OpenNF's loss-free move; recall that \name provides both
loss-freeness and order preservation.  \name's move operation takes
97$\%$ {\bf or 35X } less time (0.071ms vs 2.5ms), because, unlike
OpenNF, \name does not need to transfer state. It notifies the
datastore manager to update the relevant instance$\_$IDs.  However,
when instances are caching state, they are required to flush cached
state operations before updating instance$\_$IDs. Even then, \name
is 89\% better because it {\em flushes only operations}. 


{\bf R3: Cross-instance state sharing:} We compare \name against
OpenNF w.r.t. the performance of strongly consistent shared state
updates across NAT instances, i.e., updates are serialized according
to some global order. \figref{eval-scaling} (with 50\% load level)
shows that \name's median per-packet latency is 99\% lower than
OpenNF's (1.8$\mu$s vs 0.166ms). The OpenNF
controller receives all packets from NFs; each is forwarded to every
instance; the next packet is released only after all instances
ACK. \name's store simply serializes all instances' offloaded
operations.




\begin{table}[t]
\centering
\setlength{\tabcolsep}{0.5em}
\small
\begin{tabular}{l||cc}
& 30$\%$load& 50$\%$load\\

\hline
\hline
			
Duplicate packets & 13768 & 34351\\
Duplicate state updates & 233 & 545 \\

\end{tabular}
\compactcaption{Duplicate packet and state update at the downstream portscan detector without duplicate suppression.\label{ts:duplicate1}
}
\vspace{-0.1in}
\end{table}

{\bf R4: Chain-wide ordering:} We revisit the chain
in~\figref{fig-example-trojan}. Each scrubber instance processes
either FTP, SSH, or IRC flows. To measure the accuracy of the Trojan
detector, we added the signature of a Trojan at 11 different points in
our trace. We use three different workloads with varying
upstream NF processing speed: W1) One of the upstream NFs
adds a random delay between 50-100$\mu$s to each packet. W2)
Two of the upstreams add the random delay. W3) All three add random
delays. We observed that \name's use of chain-wide logical clocks
helps the Trojan detector identify {\bf all 11 signatures}.  We
compare against OpenNF which does not offer any chain-wide guarantees;
we find that {\bf OpenNF misses 7, 10, and 11 signatures} across
W1--W3.

{\bf R5: Duplicate suppression:} Here, we emulated a straggler NAT by
adding a random per packet delay between between 3-10$\mu$s. A portscan detector is immediately downstream from the NAT. \name launches a
clone NAT instance according to \secref{straggler}. We vary the input
traffic load.  Table~\ref{ts:duplicate1} shows the number of duplicate
packets generated by the NAT instances under different loads, as well
as the number of duplicate state updates at the portscan detector --
which happen whenever a duplicate packet triggers the scan detector to
{\em spuriously} log a connection setup/teardown attempt.  Duplicate
updates create both false positives/negatives and their incidence
worsens with load. No existing framework can detect such
duplicate updates; \name simply suppresses them.

\label{s:ft-eval}

{\bf R6: Fault Tolerance:} We study  \name failure recovery.

\begin{figure}[!t]
		\hspace{-0.4cm}%
	\begin{minipage}[b]{0.50\linewidth}
		\centering
		\includegraphics[valign=T,width=1.05\textwidth]{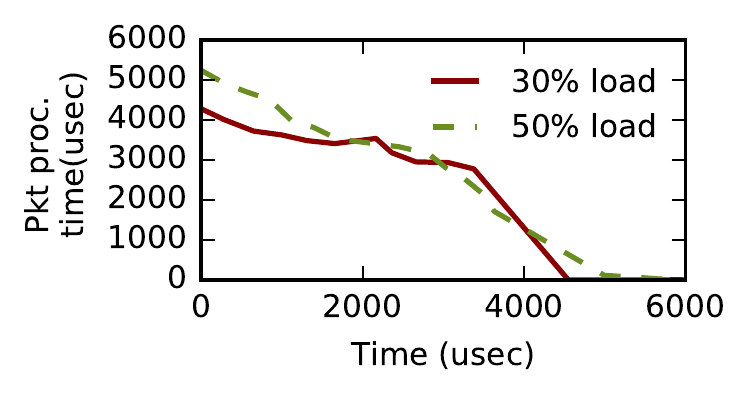}
		\vspace{-1.0em}
		\caption{Packet proc time.}
		\label{eval-ft-reply}
	\end{minipage}%
	\hfill
	\begin{minipage}[b]{0.43\linewidth} %
		\centering
		\includegraphics[valign=T, width=1.05\textwidth]{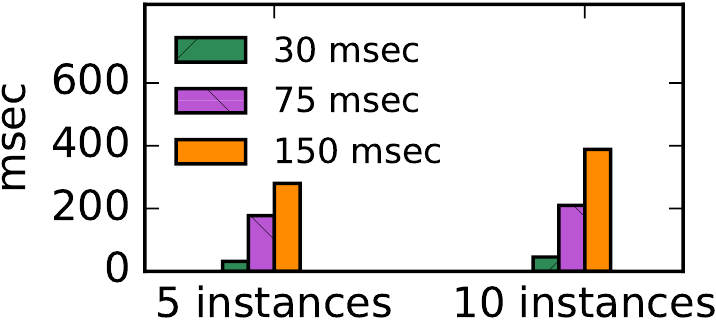}
		\vspace{0.05in}

		\caption{Store recovery.}
		\label{eval-ft-chkpt}
	\end{minipage}        
	\vspace{-0.25in}
\end{figure}

{\em NF Failure:} We fail a single NAT instance and measure the
recovery and per packet processing times. Our NAT performs
non-blocking updates without waiting for the framework ACK; here, we
use the 32bit vector (\secref{faulttolerance}) to enable recovery of
packets whose non-blocked operations are not yet committed to the store. To focus
on \name's state recovery, we assume the failover container is
launched immediately. \figref{eval-ft-reply} shows the average
processing time of packets that arrive at the new instance at two
different loads. The average is calculated over 500$\mu$s
windows. Latency during recovery spikes to over 4ms, but it only takes
{\bf 4.5ms} and {\bf 5.6ms} at 30\% and 50\% loads,
respectively, for it to return to normal.


{\em Root failure:} Recovering a root requires just
reading the last updated logical clock from the datastore and
 flow mapping from downstream NFs. This takes $<41.2\mu$s.

{\em Datastore instance failure:} Recovering a datastore instance
failure requires reading per-flow state from NFs using it, and
replaying update operations to rebuild shared state. Reading the
latest values of per-flow state is fast. Recovering shared state
however is more time-consuming.  ~\figref{eval-ft-chkpt} shows the
time to rebuild shared state with 5 and 10 NAT instances updating the
same state objects at a single store instance. We replayed the state
update operation logs generated by these instances. The instances were
processing 9.4Gbps of traffic; periodic checkpoints occurred at
intervals of 30ms, 75ms, and 150ms. The recovery time is {\bf $\le$
  388.2ms} for 10 NATs with checkpoints at 150ms intervals. In other
words, a storage instance can be quickly recovered.

	\section{Conclusion}

We presented a ground-up NFV framework called \name to support COE and
high performance for NFV chains. \name relies on managing state
external to NFs, but couples that with several caching and state
update algorithms to ensure low latency and high throughput. In
addition, it leverage simple metadata to ensure various correctness
properties are maintained even under traffic reallocation, NF
failures, as well as failures of key \name framework components.

 


	\label{EndOfPaper}
	{
		\balance
	\bibliography{ref}
	\bibliographystyle{abbrv} 
	}	
	\newpage

\appendix
\section*{Appendix}

\section{Handling non-deterministic values}
\label{non-deter}
%
%
%
%
{\bf Non-deterministic values:} Non-deterministic values, e.g.,
``gettimeofday'' or ``random'' require special handling during fault
tolerance and straggler mitigation to ensure COE. \CR{Specifically, we
require NF instances to either write every locally-computed non-deterministic value into the datastore using blocking operations or use datastore-based computation of non-deterministic values to provide determinism during fault tolerance. The datastore keeps a copy of each computed non-deterministic value. It deletes these values when their corresponding packets are deleted from the root. These stored values are used during the input replay phase to avoid divergence of internal state.}

{\bf Non-deterministic values in straggler mitigation:} When straggler
mitigation is used, to ensure that the state of an NF instance and its
clone do not diverge during straggler mitigation, \name replaces local
computation of non-deterministic values with {\em datastore based
computations}. That is, NFs request the datastore, which computes,
replies with, and stores each non-deterministic value. If a second
request for a non-deterministic value comes with the same packet
logical clock (from the straggler), the datastore emulates the
computation and returns the same value again.

\section{Proofs of Correctness and Chain Output Equivalence}
\label{proofs}
The collective action taken by all the NF instances in the chain must be equivalent of the action taken by a chain of ideal NFs. The ideal NF chain consist of NFs with infinite resources where only one instance of each type is required to handle any network load. The ideal NF processes packets in the order of their arrival. Although the ideal NF chain has infinite resources, we still assume a standard network. The network can drop or reorder packets between two {\em end hosts}.

\subsection{Consistency Guarantees of Cross-flow State Update}
\label{secproof:crossflow}
\begin{sectheorem}
\label{thm:crossflow}
	 Suppose we are given a cross-flow state $S$ and instances $\alpha$ and $\beta$ of the same NF, processing packets $P_\alpha$ and $P_\beta$, respectively. $\alpha$ and $\beta$ cannot generate a state $S''$ which is unreachable if both $P_\alpha$ and $P_\beta$ were to be processed by a single NF instance.
\end{sectheorem}

{\em Proof:} The state update corresponding to the packets $P_\alpha$ and $P_\beta$ can be applied by the store in any arbitrary order $o$. For $S''$ to be unreachable by the ideal NF, $o$ should not exist in the set of all possible orders $\{O\}$ in which packets can arrive at the ideal NF. As the network does not provide any ordering guarantees, $\{O\}$ contains all possible orders. Hence, $o$ always exists in $\{O\}$. In other words, $S$ is always reachable by an ideal NF under some particular order of updates from the input packets.

\subsection{Consistency Guarantees of Cached Cross-flow State Update}
\label{secproof:cachedcrossflow}
\begin{sectheorem}
\label{thm:cachedcrossflow}
	 Supposed we are given a cross-flow state $S$ copies of which are cached at NF instances $\alpha$ and $\beta$ of the same NF, processing packets $P_\alpha$ and $P_\beta$, respectively. $\alpha$ and $\beta$ cannot generate a state $S''$ which is unreachable if both packets are processed by a single NF instance with infinite resources. 
\end{sectheorem}

{\em Proof:} Cross-flow cached state is only used for read requests. All update operations are performed by the store. This ensures that update operations are applied on the most recent version of the state. According to \thmref{thm:crossflow}, state update always results in a consistent value, regardless of the order of the cross-flow state updates operations. Hence NF instances $\alpha$ and $\beta$  cannot generate the state $S''$.

%

\subsection{Safe Recovery of a Root Instance}
\label{secproof:rootrecovery}
\begin{sectheorem}
	\label{thm:rootrecovery} If a root with $P_i,...,P_n$ logged
	packets ($n > i > 0$) crashes after successfully sending
	$P_i,...,P_k$ packets ($i< k < n$) to downstream NF instances,
	then recovery of the root results in state $S'$ such that $S'$
	is reachable by a chain of ideal NFs under some network drop
	scenario imposed on the input traffic.
\end{sectheorem}

{\em Proof:} If the root crashes after transmitting packet $P_k$, all the remaining logged packets $P_{k+1},...,P_n$ are lost. The new root can only start from packet $P_{n+1}$. In such a case the resultant output/state is equivalent to the case of a chain of ideal NFs with $P_{k+1},...,P_n$ dropped by the network.

\subsection{Safe Recovery of an NF Instance}
\label{secproof:nfrecovery}
\begin{sectheorem}
		\label{thm:NFrecovery-base}
		If an NF instance crashes after successfully processing packet $P_{i}$,  forwarding it and pushing the corresponding state to the store, then the recovered NF instance reaches the same state as if no failure has occurred in the chain.
\end{sectheorem}

{\em Proof:} The NF crashes after successfully forwarding the state corresponding to the packet $P_i$ to the store; thus only in-transit packets are not processed ($P_{i+1}$ and onward). Since all the unprocessed in-transit packets are logged at the root, they are replayed to the new NF instance to result in the same state as if no failure has occurred.

\begin{sectheorem}
	\label{thm:NFrecovery-state}
	If an NF instance crashes after successfully processing the packet  $P_i$ and forwarding it but before the successful pushing the corresponding state update, then the recovered NF instance reaches the same state as if no failure has occurred in the chain.
\end{sectheorem}

{\em Proof:} The recovery process starts from packet $P_i$ instead of packet $P_{i+1}$ because the bit vector (\secref{faulttolerance}) value corresponding to $P_i$ at the root is non-zero which indicates that either the state has not been updated or the packet is lost. This results in replay starting from packet $P_i$. This replay regenerates the missing state update and results in the same state as if there is no failure. The duplicate suppression mechanism drops the packet $P_i$ at the message queue of the immediate downstream NF instance, preventing any duplicate packets.

\begin{sectheorem}
	\label{thm:NFrecovery-pkt}
	If an NF instance crashes after successfully processing the packet  $P_i$ and transmitting its respective state but before the successful delivery of the packet to a downstream NF, then the recovered NF instance reaches the same state as if no failure has occurred in the chain.
\end{sectheorem}

{\em Proof:} The recovery process starts from packet $P_i$ instead of packet $P_{i+1}$ because (as above) the bit vector value corresponding to $P_i$ at the root is non-zero which indicates that either the state has not been updated or the packet is lost. As the replay starts from $P_i$, the downstream NF does not experience any missing packets. The packet $P_i$ may result in a duplicate state update which is suppressed by the store. According to the \thmref{thm:NFrecovery-base}, replay of logged packets ($P_{i+1}$ and onward) recovers the state. 

\begin{sectheorem}
	\label{thm:NFrecovery-delete}
	 If the last NF instance in the chain crashes before successfully transmitting the ``delete'' request for packet $P_i$ to the root, then the recovered NF instance reaches the same state as if no failure has occurred in the chain and the end host (receiver) does not receive a duplicate packet.
\end{sectheorem}

{\em Proof:} The packet $P_i$ cannot leave the last NF without the successful transmission of ``delete'' request; thus,  the packet $P_i$ is not forwarded to the receiver. After recovery (according to \thmref{thm:NFrecovery-state} and \thmref{thm:NFrecovery-pkt}), the NF instance reaches  the same state as if no failure has occurred in the chain.

\subsection{Safe Recovery of a Store Instance}
\label{secproof:storerecovery}
\begin{sectheorem}
	\label{thm:NFrecovery-}
	If an instance of the store crashes, then each recovered per-flow state $S_P$ of each NF is equivalent to the state that would have resulted if there was no failure.
\end{sectheorem}

{\em Proof:} As $S_P$ is updated by a single NF at any given time, so
the cached value of the state at the given NF is always the most
recent state. Therefore, the new store instance reading all the cached
$S_P$s correctly recover the per-flow states.

\begin{sectheorem}
	\label{thm:case1}
	If an instance of the store crashes before any successful cross-flow state read operation, then the recovered cross-flow state 
$S_C$ must be reachable by an ideal NF under some input traffic arrival/update order. \end{sectheorem}

{\em Proof:} Write-ahead log is replayed (\secref{faulttolerance}) to reconstruct the lost cross flow state $S_C$. As none of the NFs have 
read the value, the log can be replay in any order (according to \thmref{thm:crossflow}, state will always be consistent.) to 
regenerate valid state.

\begin{sectheorem}
	If an instance of the store crashes after a successful cross-flow state read operation, then the recovered cross-flow state $S_C$ must be reachable by an ideal NF under some particular scenario.
\end{sectheorem}

{\em Proof:} As the recovery process (\secref{faulttolerance}) of cross-flow state starts from the last read value, replay of 
remaining log operations will always result in a consistent state (according to \thmref{thm:case1}.).

	\end{document}